\def \seclogo{}

\documentclass{ieeetj}
\usepackage{hyperref,booktabs,amsmath,adjustbox,placeins}
\hypersetup{
    colorlinks=true,
    citecolor=blue,
    linkcolor=blue,
    filecolor=magenta,      
    urlcolor=blue,
}

\begin{document}

\title{ICASSP 2023 Speech Signal Improvement Challenge}
\author{Ross Cutler, Ando Saabas, Babak Naderi, Nicolae-C\u{a}t\u{a}lin Ristea, Sebastian Braun, Solomiya Branets}
\affil{Microsoft Corporation, Redmond, USA}
\corresp{Corresponding author: Ross Cutler (email:ross.cutler@microsoft.com).}

\begin{abstract}
The ICASSP 2023 Speech Signal Improvement Challenge is intended to stimulate research in the area of improving the speech signal quality in communication systems. The speech signal quality can be measured with SIG in ITU-T P.835 and is still a top issue in audio communication and conferencing systems. For example, in the ICASSP 2022 Deep Noise Suppression challenge, the improvement in the background and overall quality is impressive, but the improvement in the speech signal is not statistically significant. To improve the speech signal the following speech impairment areas must be addressed: coloration, discontinuity, loudness, reverberation, and noise. A training and test set was provided for the challenge, and the winners were determined using an extended crowdsourced implementation of ITU-T P.804's listening phase. The results show significant improvement was made across all measured dimensions of speech quality. 
\end{abstract}


\maketitle

\section{Introduction}
Audio telecommunication systems such as remote collaboration systems (Microsoft Teams, Skype, Zoom, etc.), smartphones, and telephones are used by nearly everyone on the planet and have become essential tools for both work and personal usage. Since the invention of the telephone in 1876 by Alexander Graham Bell, audio engineers, and researchers have innovated to improve the speech quality of telecommunication systems, with the ultimate goal of making audio telecommunication systems as good or better than face-to-face communication.  After nearly 150 years of effort, there is still a long way to go toward this goal, especially with the use of mainstream devices. For example, it is still common to hear frequency response distortions, isolated and non-stationary distortions, loudness issues, reverberation, and background noise in audio calls.

The ICASSP 2023 Speech Signal Improvement Challenge is intended to stimulate research in the area of improving the send speech signal\footnote{In telecommunication, the audio captured by a near end microphone, processed, and sent to the far end is called the send signal.} quality in mainstream telecommunication systems. Subjective speech quality assessment is the gold standard for evaluating speech enhancement, processing, and telecommunication systems. The ITU-T has developed several recommendations for subjective speech quality assessment. In particular, the ITU-T Rec.~P.835 \cite{noauthor_itu-t_2003} provides a lab-based subjective evaluation framework targeting systems that include noise suppression algorithm that gives quality scores of the speech signal (SIG), background noise (BAK), and overall quality (OVRL). In this framework, participants are asked to listen to short clips of speech in a controlled environment and rate the quality of each clip in terms of the speech signal, background noise, and overall quality on three discrete Likert scales (where 1 is Bad quality and 5 is Excellent quality). Each clip is measured by multiple raters, and the results are averaged to obtain a Mean Opinion Score (MOS). By measuring SIG, BAK, and OVRL, P.835 provides a more reliable subjective assessment \cite{noauthor_itu-t_2003} and allows researchers to determine which area to focus on for improving the overall quality.

The speech signal is still a top issue in audio telecommunication and conferencing systems. For example, in the ICASSP 2022 Deep Noise Suppression Challenge \cite{dubey_icassp_2022}, the improvement in BAK and OVRL quality is impressive, but no improvement in SIG was observed. The same was true for the INTERSPEECH 2021 Deep Noise Suppression Challenge \cite{reddy_interspeech_2021}, and for the more recent ICASSP 2023 Deep Noise Suppression Challenge \cite{dubey_icassp_2023}, which focuses on personalized noise suppression. Table \ref{tab:improvement} shows the amount of improvement in SIG, BAK, and OVRL to get excellent quality rated speech (MOS=5) for the ICASSP 2022 Deep Noise Suppression Challenge. This shows the key area of improvement is SIG, which has 2.3$\times$ more improvement opportunities than BAK. To improve SIG, the following dimensions of speech quality should be improved \cite{noauthor_itu-t_2017-1}: 

\begin{itemize}
    \item Coloration: Frequency response distortions
    \item Discontinuity: Isolated and non-stationary distortions
    \item Loudness: Important for the overall quality and intelligibility
    \item Reverberation: Room reverberation of speech and noise signals
    \item Noisiness: Background noise and circuit and coding noise
\end{itemize}

The correlation of SIG to these dimensions is given in Figure \ref{fig:p804_corr}. Theoretically, improving BAK is not necessary to improve SIG as they are orthogonal metrics by design. However, in practice, it is hard for subjective test participants to assess speech signal quality in the presence of strong dominant background noise.

\section{Related work}
While there have been previous challenges in background noise and reverberation, there have been no challenges in coloration and loudness and a limited challenge in discontinuities (see Table~\ref{tab:related}). Moreover, there have been no previous challenges that explicitly measure and target improving SIG.

There are many previous methods to improve noisiness, coloration, discontinuity, loudness, and reverberation separately. Two new methods that target universal improvement of the speech signal are \cite{su_hifi-gan-2_2021,serra_universal_2022}. 

The ITU-T has developed several recommendations for subjective speech quality assessment. ITU-T P.800 \cite{noauthor_itu-t_1996}  describes lab-based methods for the subjective determination of speech quality, including the Absolute Category Rating (ACR). ITU-T P.808 \cite{noauthor_itu-t_2018} describes a crowdsourcing approach for conducting subjective evaluations of speech quality. It provides guidance on test material, experimental design, and a procedure for conducting listening tests in the crowd. The methods are complementary to laboratory-based evaluations described in P.800. An open-source implementation of P.808 is described in \cite{naderi_open_2020}. An open-source implementation of P.835 is described in \cite{naderi_subjective_2021}. More recent multidimensional speech quality assessment standards are ITU-T P.863.2 \cite{noauthor_itu-t_2022} and P.804 \cite{noauthor_itu-t_2017-1} (listening phase), which measure perceptual dimensions of speech quality namely noisiness, coloration, discontinuity, and loudness (see Table \ref{tab:evaluation}).

Intrusive objective speech quality assessment tools such as Perceptual Evaluation of Speech Quality (PESQ) \cite{rix_perceptual_2001} and Perceptual Objective Listening Quality Analysis (POLQA) \cite{beerends_perceptual_2013} require a clean reference of speech. Non-intrusive objective speech quality assessment tools like ITU-T P.563 \cite{noauthor_itu-t_2011} do not require a reference, though it has a low correlation to subjective quality \cite{avila_non-intrusive_2019}. Newer neural net-based methods such as \cite{avila_non-intrusive_2019, reddy_dnsmos_2021, reddy_dnsmos_2022, yi_conferencingspeech_2022} provide better correlations to subjective quality. NISQA \cite{mittag_nisqa_2021} is an objective metric for P.804, though the correlation to subjective quality is not sufficient to use as a challenge metric (in the ConferencingSpeech 2022 Challenge \cite{yi_conferencingspeech_2022} NISQA was used as a baseline model and achieved a Pearson Correlation Coefficient = 0.724 to MOS).

\section{Challenge description}
This challenge benchmarks the performance of speech enhancement models with a real (not simulated) test set. The telecommunication scenario is the near end only send signal; it does not include echo impairments (there is no far end speech or noise). Participants evaluated their speech enhancement model (SEM) on a test set and submitted the results (clips) for subjective evaluation. 

\subsection{Challenge tracks}
The challenge has two tracks:

\begin{enumerate}
    \item Real-time SEM
    \item Non-real-time SEM
\end{enumerate}

The goal of the first track is to develop something that can be used today on a typical personal computer, while the goal of the second track is to develop something that could be run on computers much faster than a typical personal computer or be run offline. 

\subsection{Latency and runtime requirements}
Algorithmic latency is defined by the offset introduced by the whole processing chain including short-time Fourier transform (STFT), inverse STFT, overlap-add, additional lookahead frames, etc., compared to just passing the signal through without modification. It does not include buffering latency. Some examples are:

\begin{itemize}
    \item A STFT-based processing with window length = 20 ms and hop length = 10 ms introduces an algorithmic delay of window length – hop length = 10 ms.
    \item A STFT-based processing with window length = 32 ms and hop length = 8 ms introduces an algorithmic delay of window length – hop length = 24 ms.
    \item An overlap-save-based processing algorithm introduces no additional algorithmic latency.
    \item A time-domain convolution with a filter kernel size = 16 samples introduces an algorithmic latency of kernel size – 1 = 15 samples. Using one-sided padding, the operation can be made fully “causal”, i.e., left-sided padding with kernel size - 1 samples would result in no algorithmic latency.
    \item A STFT-based processing with window\_length = 20 ms and hop\_length = 10 ms using 2 future frames information introduces an algorithmic latency of (window\_length – hop\_length) + 2 * hop\_length = 30 ms.
\end{itemize}

Buffering latency is defined as the latency introduced by block-wise processing, often referred to as hop length, frame-shift, or temporal stride. Some examples are:

\begin{itemize}
    \item A STFT-based processing has a buffering latency corresponding to the hop size.
    \item A overlap-save processing has a buffering latency corresponding to the frame size.
    \item A time-domain convolution with stride 1 introduces a buffering latency of 1 sample.
\end{itemize}

Real-time factor (RTF) is defined as the fraction of time it takes to execute one processing step. For an STFT-based algorithm, one processing step is the hop size. For a time-domain convolution, one processing step is 1 sample. RTF = compute time / time step.

All models submitted to this challenge must meet all of the below requirements:

\begin{enumerate}
    \item To be able to execute an algorithm in real-time, and to accommodate for variance in compute time which occurs in practice, we require RTF $\leq$ 0.5 in the challenge on an Intel Core i5 Quadcore clocked at 2.4 GHz using a single thread.
    \item Algorithmic latency + buffering latency $\leq$ 20 ms.
    \item No future information can be used during model inference.
\end{enumerate}

\begin{table}
    \centering
    \caption{Amount of improvement (in Differential MOS (DMOS)) remaining to get excellent quality (MOS=5) rated speech based on the ICASSP 2022 DNS Challenge \cite{dubey_icassp_2022}. DMOS is on a scale of 0-4, where 0 is no difference and 4 is very annoying compared to excellent quality speech.}
    \begin{tabular}{c c}
        \toprule
        Area & Headroom (DMOS) \\
        \midrule
        SIG & 0.70 \\
        BAK & 0.30 \\
        OVRL & 0.87 \\
        \bottomrule
    \end{tabular}
    
    \label{tab:improvement}
\end{table}

\begin{table}
    \centering
    \caption{Related challenges.}
    \begin{tabular}{l l}
    \toprule
        Area & Related challenge \\
        \midrule
        Noisiness & Deep Noise Suppression \cite{reddy_interspeech_2020,reddy_icassp_2021,reddy_interspeech_2021,dubey_icassp_2022,dubey_icassp_2023} \\
        Coloration & None \\
        Discontinuity & Packet Loss Concealment \cite{diener_interspeech_2022} \\
        Loudness & None \\
        Reverberation & REVERB \cite{kinoshita_summary_2016} \\
        Echo & AEC Challenge \cite{sridhar_icassp_2021,cutler_interspeech_2021,cutler_icassp_2022,cutler_icassp_2023}\\
        \bottomrule
    \end{tabular}
    
    \label{tab:related}
\end{table}

More details of the challenge are available at \url{https://aka.ms/sig-challenge}.
\section{Training set}
This challenge suggested using the ICASSP 2022 Deep Noise Suppression Challenge \cite{dubey_icassp_2022}  and ICASSP 2022 Acoustic Echo Cancellation Challenge \cite{cutler_icassp_2022} training and test sets for training. The AEC Challenge training set in particular includes over 10K unique environments, devices, and speakers. The near end single talk clips have been rated using P.835 and are provided, which can be used during training to improve SIG and OVRL. 

However, any training set could have been used, such as \cite{kinoshita_summary_2016}, \cite{li_dds_2022}, \cite{mysore_can_2015}.

\section{Test set}
\label{sec:test_set}
\begin{figure*}
    \centering
    \includegraphics[width=1.5\columnwidth]{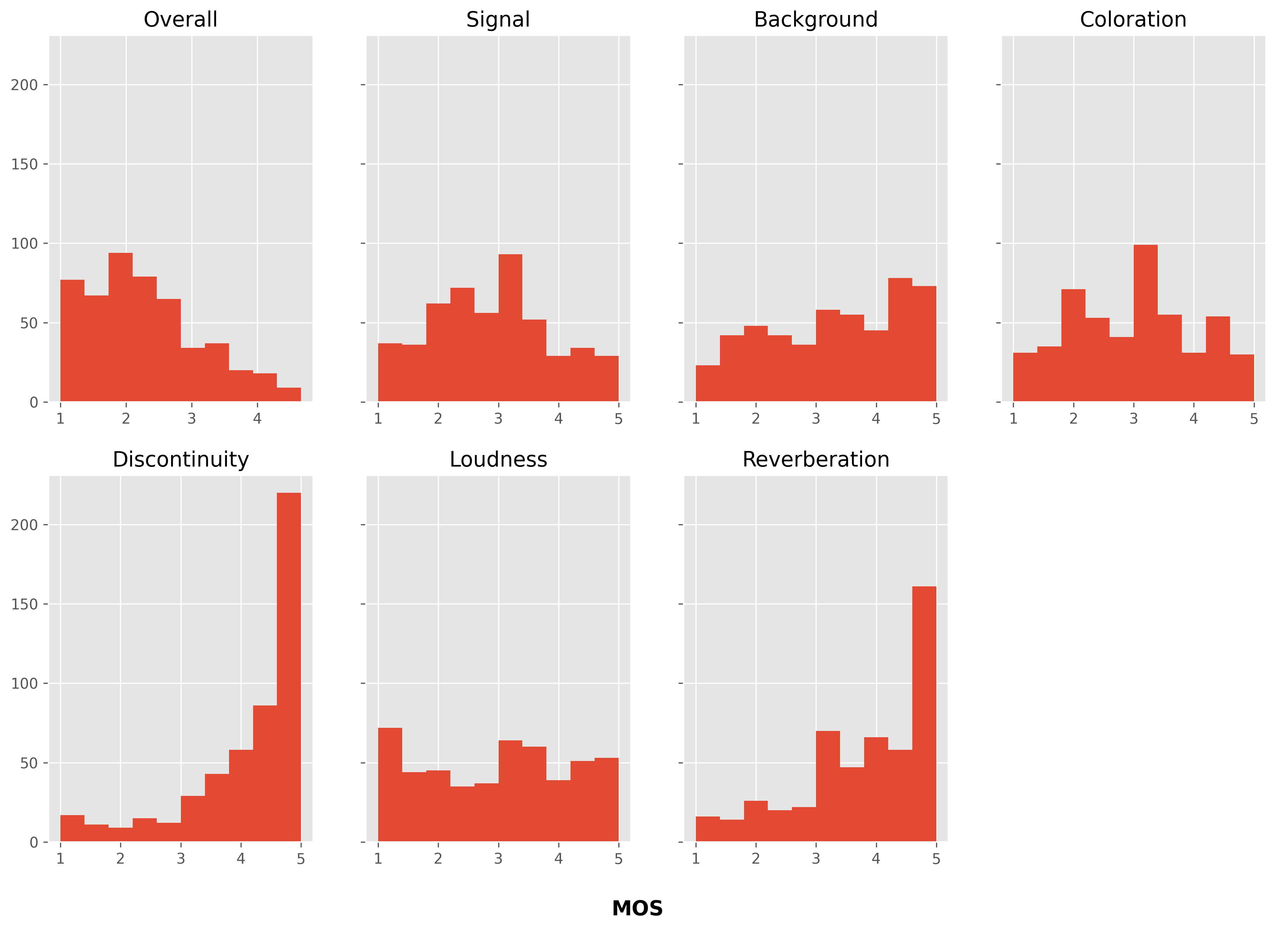}
    \caption{Distribution of subjective scores of clips in the blind test set. Ideally, the distribution would be uniform for each dimension, but it is skewed for discontinuity and reverberation.}
    \label{fig:noisy_dis}
\end{figure*}

The test set consisted of 500 send clips, each using a unique device, environment, and person speaking. The clips were captured from both PCs and mobile devices using the same methodology as described in \cite{cutler_icassp_2022}. The recordings were stratified to have an approximately uniform distribution for the impairment areas listed in Table \ref{tab:evaluation}. The test set language contains English, German, Dutch, French, and Spanish languages, with the majority of files (around 80\%) in English. The test set was released near the end of the competition. The distribution of subjective ratings based on P.804 (see Section \ref{sec:evaluation}) of the test set for all dimensions is shown in Figure \ref{fig:noisy_dis}.

\section{Evaluation methodology}
\label{sec:evaluation}
The challenge evaluation is based on a subjective listening test. We have developed an extension of P.804 (listening phase) / P.863.2 (Annex A) based on crowdsourcing and the P.808 toolkit \cite{naderi_open_2020} for subjective evaluation. In particular, we added reverberation, speech quality, and overall quality to  P.804's listening phase (see Table \ref{tab:evaluation}). Details of this P.804 extension are given in Section \ref{sec:framework} and \cite{naderi_multi-dimensional_2023}.

The challenge metric $M\in[0,1] $ is:

\begin{equation}
    M = \frac{(\text{SIG}-1)/4 + (\text{OVRL}-1)/4}{2}
\end{equation}

\noindent In addition, differential SIG (DSIG) must be $> 0$. Since OVRL $\sim$ BAK + SIG \cite{naderi_subjective_2021}, BAK should also be improved to increase OVRL performance. The other metrics measured in Table \ref{tab:evaluation} are informational only. However, in Section \ref{sec:model} we show that overall is influenced by the speech signal, reverberation, loudness, discontinuity, coloration, and noisiness, so optimizing each is a good strategy. The clips are evaluated as fullband (48 kHz) in P.804, so frequency extension can help. 

\begin{table*}
    \caption{Speech quality areas from P.804 listening phase (the first four) plus three additional areas.}
    \label{tab:evaluation}
    \setlength\tabcolsep{2.0pt}
    \centering
    \begin{tabular}{c c c }
        \toprule
        Area & Description & Possible source \\
        \midrule
        Noisiness & Background noise, circuit noise, coding noise; BAK & Coding, circuit or background noise; device \\
        Coloration & Frequency response distortions & Bandwidth limitation, resonances, unbalanced freq. response \\
        Discontinuity & Isolated and non-stationary distortions & Packet loss; processing; non-linearities \\
        Loudness & Important for the overall quality and intelligibility & Automatic gain control; mic distance \\
        Reverberation & Room reverberation of speech and noise & Rooms with high reverberation \\
        Speech Signal & SIG & \\
        Overall & OVRL & \\
     \bottomrule
    \end{tabular}
\end{table*}

\subsection{Online subjective evaluation framework}
\label{sec:framework}
We extended the P.808 Toolkit \cite{naderi_open_2020} to include a test template for a multi-dimensional quality assessment. The toolkit provides scripts for preparing the test, including packing the test clips in small test packages, preparing the reliability check questions, and analyzing the results. We ask participants to rate the perceptual quality dimensions of speech namely coloration, discontinuity, noisiness, loudness, reverberation, signal quality, and overall quality of each audio clip. In the following, each section of the test template, as seen by participants, is described. These sections are predefined and only the audio clips under the test will be changed from one study to another.

In the first section, the participant’s eligibility and device suitability are tested and a qualification is assigned to those that pass which remains valid for the entire experiment. The participant's hearing ability is evaluated through digit-triplet-test~\cite{naderi_towards_2020}. Moreover, we test if their listening device supports the required bandwidths (i.e., fullband, wideband, and narrowband); details are in Section~\ref{optimization}).

\begin{figure}
    \centering
  \includegraphics[width = 0.8\columnwidth]{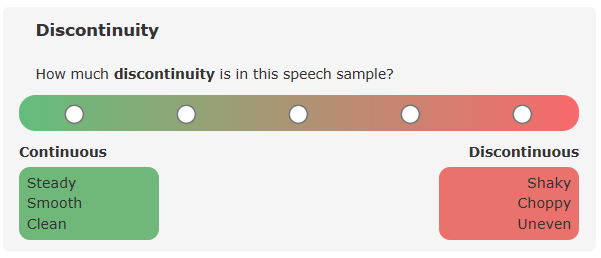}
  \caption{Sub-dimensions are rated on a 5-point discrete scale with descriptive adjectives on poles.}
  \label{fig:scale}
\end{figure}

Next, the participant's environment and device are tested using a modified-JND test~\cite{naderi_application_2020} in which they should select which stimulus from a pair has a better quality in four questions. A temporal certificate will be issued for participants after passing this section which expires after two hours and consequently repeating this section will be required. Detailed instructions are given in the next section including introducing the rating scales and providing multiple samples for each perceptual dimension. Participants are required to listen to all samples for the first time. Figure~\ref{fig:scale} illustrates how the rating scale for quality dimensions is presented to participants. In addition, we used a Likert 5-point scale for signal quality and overall quality as specified by ITU-T Rec.~P.835. In the Training section participants should first adjust the playback loudness to a comfortable level by listening to a provided sample and then rate 7 audio clips. This section is similar to the ratings section, but the platform provides live feedback based on their ratings. By completing this section a temporal certificate is assigned to the participants which is valid for one hour. Last is the Ratings section, where participants listen to ten audio clips and two gold standard and trapping questions and cast their votes on each scale. The gold standard questions are the ones that the experimenter already knows their answers (being excellent or bad) and participants are expected to vote on each scale with a minor deviation from known the answer~\cite{naderi_towards_2020}.  Trapping questions are questions in which a synthetic voice is overlaid to a normal clip and asks participants to provide a specific vote to show their attention \cite{naderi_effect_2015}. For this test, we provide scripts for creating the trapping clips, which ask participants to select answers reflecting the best or worst quality in all scales. For rating an audio clip, the participant should first listen to the end of the clip, and then they start casting their votes. During that time, the audio will be played back in a loop. After participants finish with a test set, they can continue with the next one where only the rating section will be shown until other temporal certificates are valid. By the expiration of any certificate, the corresponding section will be shown when they start the next test set.

\begin{table*}
    \centering
     \caption{Labels on each scale's pole and descriptive adjectives provided to participants. Terms used in ITU-T Rec.~P.804 are marked in red.}
  \label{tab:terms}
  \includegraphics[width = 1.4\columnwidth]{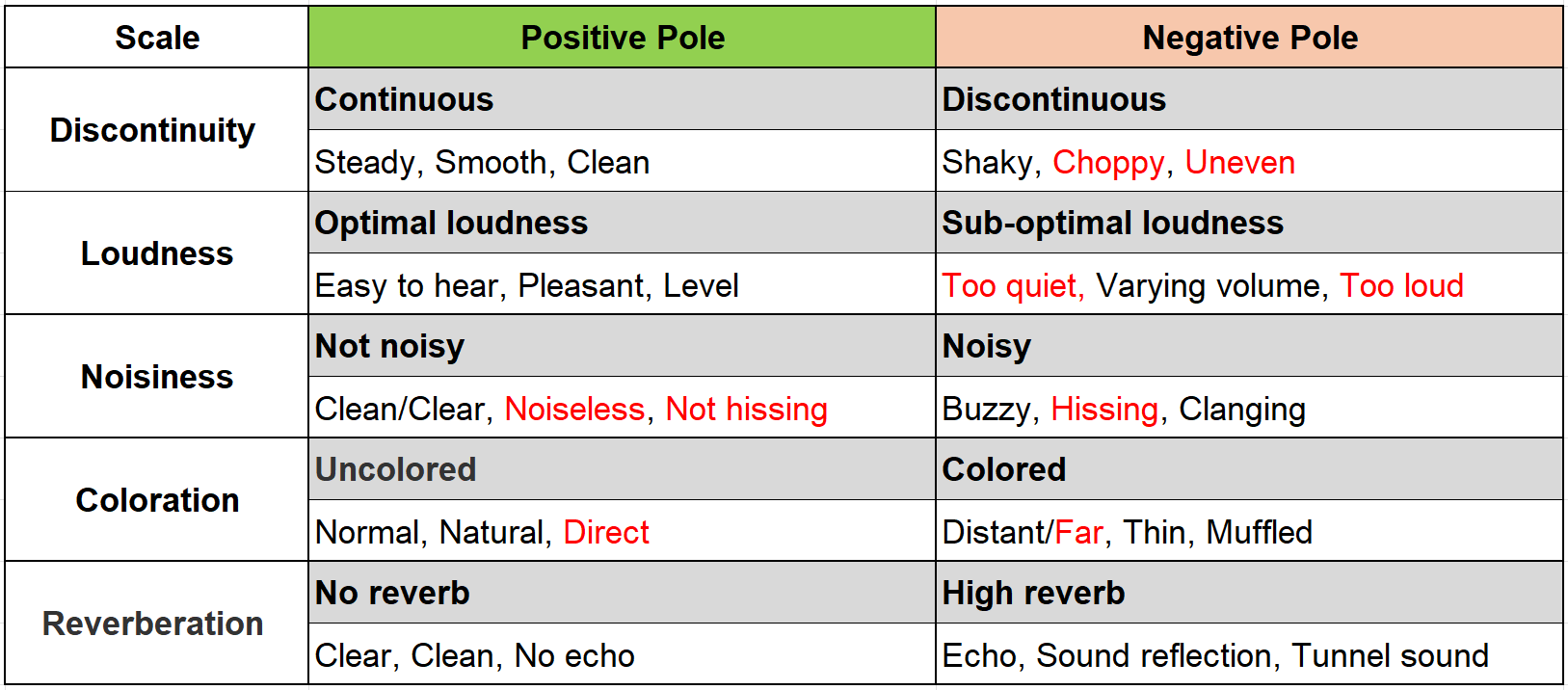}
\end{table*}

\subsubsection{Survey optimization}
\label{optimization}
We utilized the multi-scale template in various research studies and improved it through the incorporation of experts and test participant feedback.

\textbf{Descriptive adjectives:} The understanding of perceptual dimensions might not be intuitive for naive test participants, therefore the P.804 recommendation includes a set of descriptive adjectives to describe the presence or absence of each quality dimension. We expanded this list through multiple preliminary studies, where participants were asked to listen to samples from each perceptual dimension and name three adjectives that best describe them. For each dimension, we selected the top three most frequently selected terms and presented them below each pole of the scale, as shown in Figure~\ref{fig:scale}. The list of selected terms is reported in Table~\ref{tab:terms}. We used discrete scales for dimensions to be consistent with signal and overall scales. 

\textbf{Bandwidth check:} This test ensures the participant devices support the expected bandwidth. The test consists of five samples, and each has two parts separated by a beep tone. The second part is the same as the first part but in three samples superimposed by additive noise. Participants should listen to each sample and select if both parts have the same or different quality. We filtered the white noise with the following bandpass filters: 3.5-22K (all devices should play the noise), 9.5-22k (super-wideband or fullband is supported), and 15-22K (fullband is supported).

\textbf{Gold questions:}  Gold questions are widely used in crowdsourcing  ~\cite{naderi_towards_2020}. Here we observed gold questions that represent the strong presence of an impairment on one dimension and the clear absence of impairment on all dimensions can best reveal an inattentive participant. 

\textbf{Randomization:} We randomize the presentation order of scales for each participant. However, the signal and overall quality are always presented at the end. The randomized order is kept for each participant until a new round of training is required.

\section{Results}
There were 7 entries for the real-time track and 5 for the non-real-time track, though the top 3 for non-real-time track were identical submissions to the real-time track and therefore were only considered for the real-time track. Team Cvt-tencent was statistically tied with Legends-tencent and withdrew. 

The P.804 and P.835 subjective results for both tracks are given in Table \ref{fig:track1_results} and Table \ref{fig:track2_results}. The ANOVAs for each track are given in Table \ref{tab:track1_anova} and Table \ref{tab:track2_anova}. P.835 results are given for reference only but agree with the P.804 results. Objective results are given in Table \ref{fig:objective_results}.

\begin{table*}
    \centering
     \caption{Real-time track challenge results.}
    \label{fig:track1_results}

    \includegraphics[width=\textwidth]{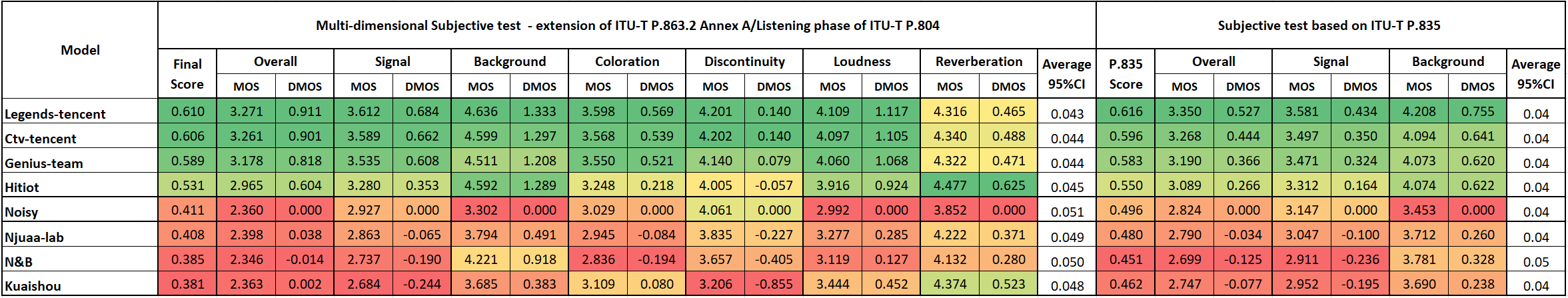}
\end{table*}

\begin{table*}
    \centering
     \caption{Non-real-time track challenge results. Teams with a * had identical submissions to the real-time track.}
    \label{fig:track2_results}

    \includegraphics[width=\textwidth]{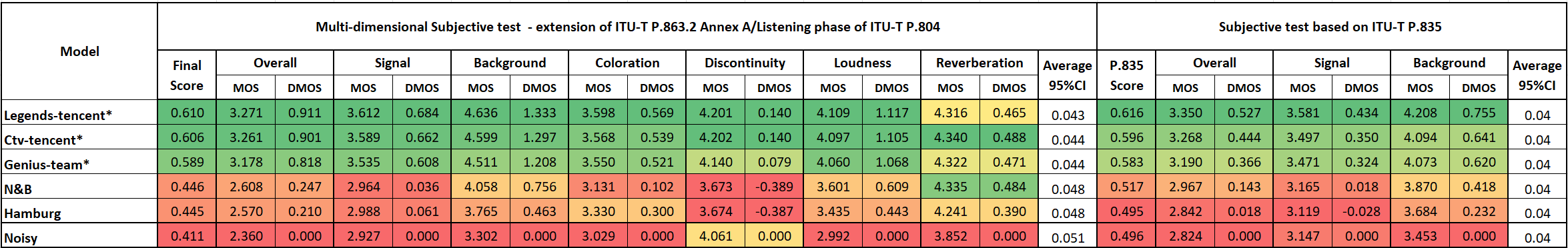}
\end{table*}

\begin{table*}
\centering
\caption{Real-time track ANOVA. The pairwise p-values are shown for the lower-triangular matrix.}
\label{tab:track1_anova}
\begin{tabular}{cccccccc}
\toprule
Team & Legends-tencent & Ctv-tencent & Genius-team & Hitiot & Noisy & Njuaa-lab & N\&B  \\
\midrule
Ctv-tencent & 0.609 &  & & & & & \\
Genius-team & 0.001 & 0.004 & & & & & \\
Hitiot & 0.000 & 0.000 & 0.000 & & & & \\
Noisy  & 0.000 & 0.000 & 0.000 & 0.000 & & & \\
Njuaa-lab & 0.000 & 0.000 & 0.000 & 0.000 & 0.681 & & \\
N\&B  & 0.000 & 0.000 & 0.000 & 0.000 & 0.000 & 0.002 & \\
Kuaishou  & 0.000 & 0.000 & 0.000 & 0.000 & 0.000 & 0.000 & 0.094 \\
\bottomrule
\end{tabular}
\end{table*}

\begin{table*}
\centering
\caption{Non-real-time track ANOVA. The pairwise p-values are shown for the lower-triangular matrix.}
\label{tab:track2_anova}
\begin{tabular}{cccccc}
\toprule
Team & Legends-tencent & Ctv-tencent & Genius-team & N\&B & Hamburg   \\
\midrule
Ctv-tencent & 0.609 & & & &  \\
Genius-team & 0.001 & 0.004 & & &  \\
N\&B  & 0.000 & 0.000 & 0.000 & &   \\
Hamburg & 0.000 & 0.000 & 0.000 & 0.285 & \\
Noisy & 0.000 & 0.000 & 0.000 & 0.000 & 0.000  \\
\bottomrule
\end{tabular}
\end{table*}

\begin{table*}
    \centering
     \caption{The objective results on the blind set obtained with DNSMOS model \cite{reddy_dnsmos_2022} (MOS$\_$SIG, MOS$\_$BAK, MOS$\_$OVR), and NISQA \cite{mittag_nisqa_2021} (NISQA$\_$MOS etc.).}
    \label{fig:objective_results}

    \includegraphics[width=0.8\textwidth]{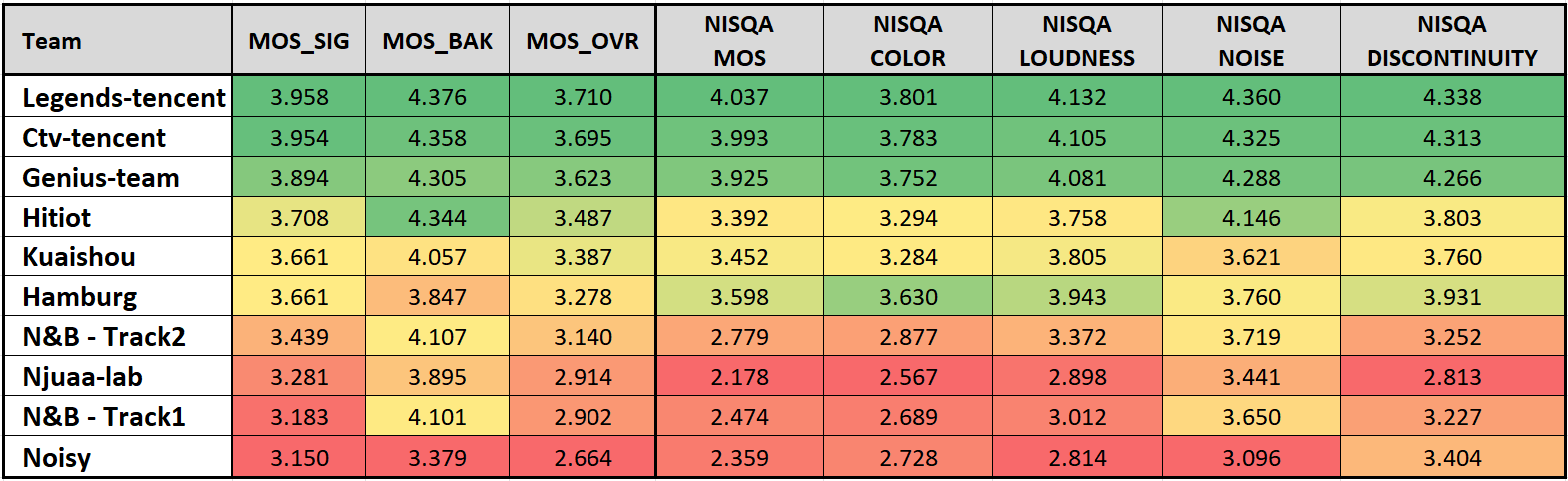}
\end{table*}

\section{Analysis}

\subsection{Comparison of methods}
A high-level comparison of the top-5 entries is given in Table \ref{tab:top5} and Table \ref{tab:models}. Some observations are given below:

\begin{table*}
\centering
\caption{Comparison of the top five teams} 
\label{tab:top5}
\setlength\tabcolsep{4.0pt}
\begin{tabular}{c c r r c r c c c c} 
\toprule
Place & Track & Team & Params & Real-time & Training set & Training set & Stages & Domain & $M$ \\
& & & & factor & & hours & & & \\
\midrule
1 & Real-time & Legends-tencent \cite{chen_gesper_2023} & 12.1 M & 0.37  & DNS \cite{dubey_icassp_2022}, private & 1500 & 3 & time, STFT & 0.610 \\ 
2 & Real-time & Genius-team \cite{zhu_ssi-net_2023} & 5.2 M & 0.36  & DNS \cite{dubey_icassp_2022} & 1500 & 2 & time, STFT & 0.589 \\ 
3 & Real-time & HITIoT \cite{zhang_half-temporal_2023} & 9.2 M & 0.36 & DNS \cite{dubey_icassp_2022} & 1500 & 1 & STFT & 0.531 \\
4 & Non-real-time & N\&B \cite{liu_two-stage_2023} & 10 M & 1.48 & DNS \cite{dubey_icassp_2022} & 421 & 2 & STFT & 0.446 \\
5 & Non-real-time & Hamburg \cite{richter_speech_2023} & 55.7 M & 30.1 & VCTK \cite{yamagishi_cstr_2019} & 28.2 & 1 & STFT & 0.445 \\
\midrule
PCC to $M$ &  &  & -0.58 & -0.60 & & 0.91 & 0.61 & &  \\
\bottomrule
\end{tabular}
\end{table*}

\begin{table*}
\centering
\caption{Models used by the top five teams} 
\label{tab:models}
\begin{tabular}{l l} 
\toprule
Team & Model \\
\midrule
Legends-tencent \cite{chen_gesper_2023} & AGC $\rightarrow$ GSM-GAN (Restore) $\rightarrow$ Enhance \\ 
Genius-team \cite{zhu_ssi-net_2023} & TRGAN (Restore) $\rightarrow$ MTFAA-Lite (Enhance) \\ 
Hitot \cite{zhang_half-temporal_2023} & Half temporal, half frequency attention U-Net \\
N\&B \cite{liu_two-stage_2023} & GateDCCRN (Repairing) $\rightarrow$ GateDCCRN, S-DCCRN (Denoising) \\
Hamburg \cite{richter_speech_2023} & Generative diffusion model (modified NCSN++) \\
\bottomrule
\end{tabular}
\end{table*}

\begin{figure*}
    \centering
    \includegraphics[width=1.5\columnwidth]{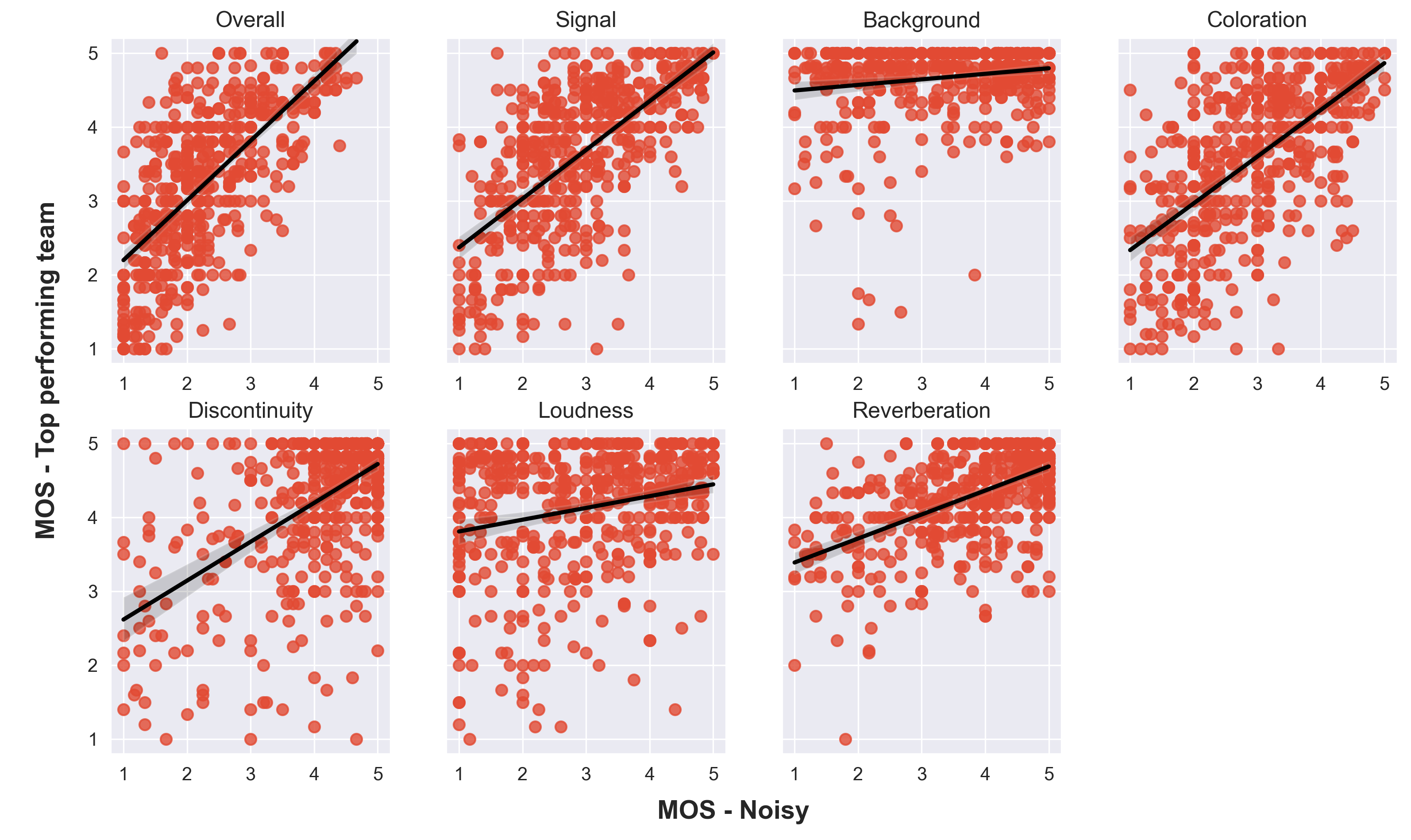}
    \caption{Distribution of subjective ratings before (X-axis) and after applying the winning model by Legends-tencent \cite{chen_gesper_2023} (Y-axis). Each dot is a single audio clip, and a best-fit line is shown. No processing would be a diagonal line from (1,1) to (5,5). Background is close to ideal, while loudness degrades excellent loudness (MOS=5) inputs.}
    \label{fig:top_model}
\end{figure*}

\begin{itemize}
    \item The top entries improved SIG by DMOS $> 0.6$, unlike previous DNS challenges which had no SIG improvement \cite{dubey_icassp_2022,reddy_interspeech_2021}.  
    \item The correlation between the training set hours (the total duration of data used) and the overall score is PCC = 0.91. The models with larger training sets tended to do better.
    \item The correlation between the runtime factor and the overall score is PCC = -0.60. We expected the non-real-time track entries to exceed the performance of the real-time track, but that was not the case. We observed a similar fact in the INTERSPEECH 2021 Deep Noise Suppression Challenge \cite{reddy_interspeech_2021}, where the non-real-time track also performed significantly worse than the real-time track. In both cases, we received more entries in the real-time track than non-real-time track, and there may be more researchers working on real-time speech enhancement than non-real-time speech enhancement. One approach to get better non-real-time models is to take the winner of the real-time track and increase the model size and complexity by 100x, very likely increasing the performance while making it no longer real-time. 
    \item The correlation between the model size and the overall score is PCC = -0.58. Smaller models tended to perform better.
    \item The correlation between the number of stages and the overall score is PCC = 0.61. More stages tended to perform better.
    \item The top model by team Legends-tencent \cite{chen_gesper_2023} significantly improved all measured speech quality dimensions, and did the best in all dimensions except reverberation. Their performance is illustrated in Figure~\ref{fig:top_model}. 
    \item A successful strategy used by teams Legends-tencent \cite{chen_gesper_2023}, Genius-team \cite{zhu_ssi-net_2023}, and HITIoT \cite{liu_two-stage_2023} is a restoration module followed by a speech enhancement module. The generative models for restoration by teams Legends-tencent \cite{chen_gesper_2023} and Genius-team \cite{zhu_ssi-net_2023} perform particularly well.
    \item There is still significant room for improvement in this test set for OVRL and SIG.
    \item None of the teams used the ICASSP 2022 Acoustic Echo Cancellation Challenge \cite{cutler_icassp_2022} dataset for training, even though it has thousands of clips of real-world speech signal impairments. This is likely because there is no clean speech available for this dataset, and using it would require semi-supervised or unsupervised training. Rather, all teams used the ICASSP 2022 Deep Noise Suppression Challenge \cite{dubey_icassp_2022} for a training set, and the winning team Legends-tencent \cite{chen_gesper_2023} augmented that with a private training set. 
\end{itemize}

\begin{figure*}
    \centering
    \includegraphics[width=1.25\columnwidth]{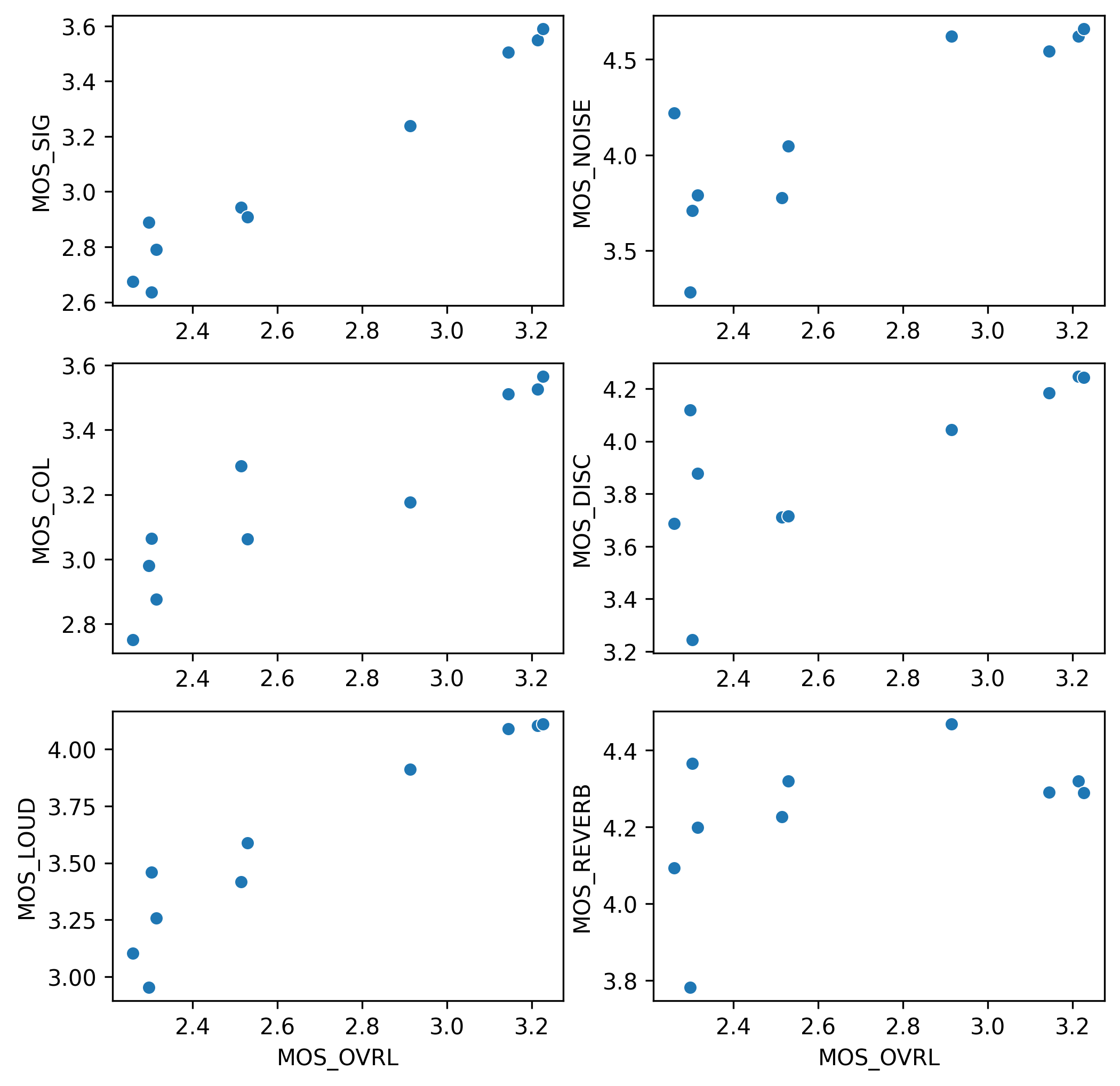}
    \caption{Distribution of subjective test dimensions for all entries in model level.}
    \label{fig:distribution}
\end{figure*}

\begin{figure}
    \centering
    \includegraphics[width=0.99\columnwidth]{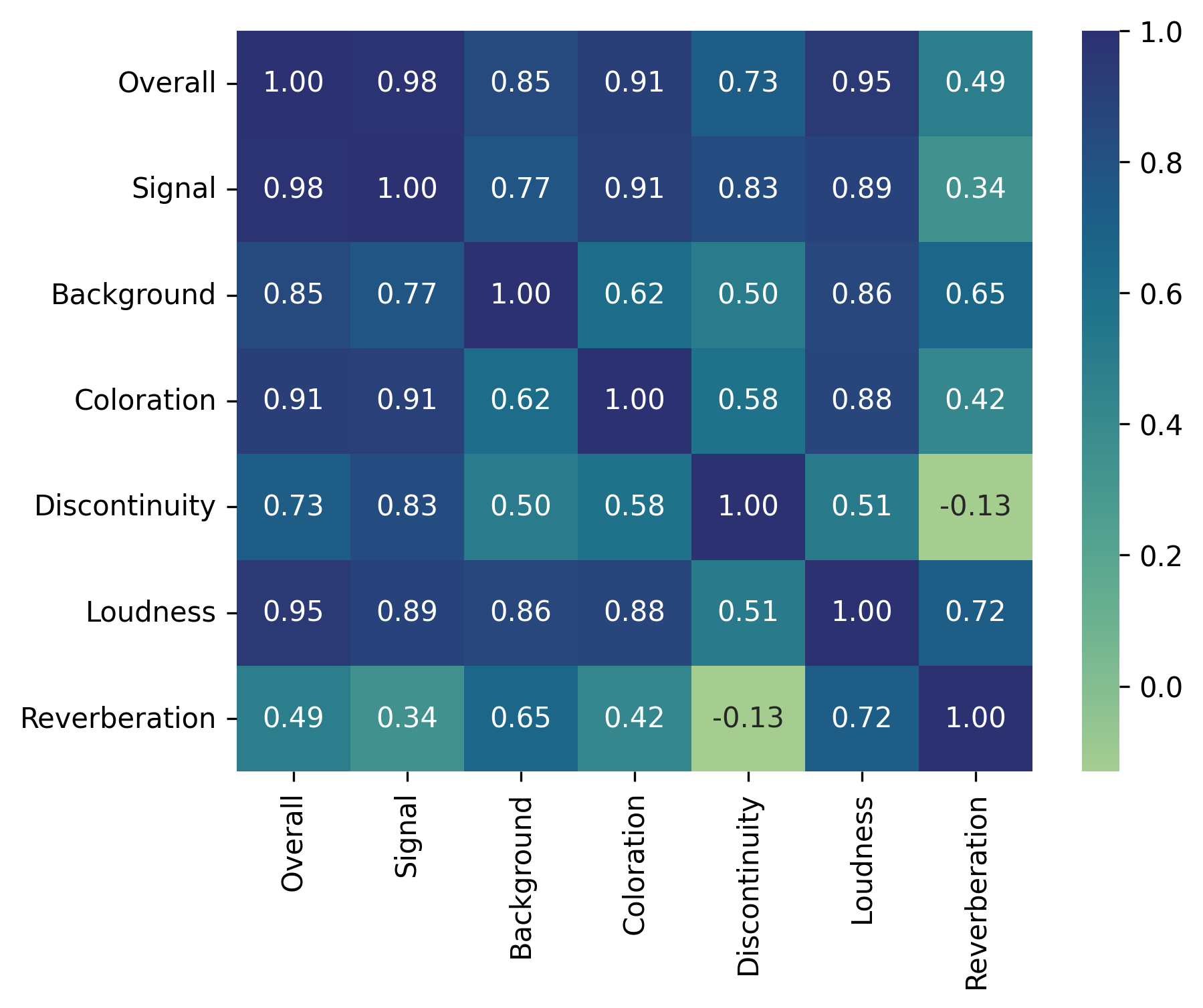}
    \caption{Pearson correlation between different subjective test dimensions for all entries in model level.}
    \label{fig:p804_corr}
\end{figure}

\subsection{Distribution of dimensions}
Figure \ref{fig:distribution} shows the distribution of the subjective dimensions compared to overall quality at the model level. All of the dimensions except discontinuity and reverberation have a significant linear correlation to the overall quality (see Figure \ref{fig:p804_corr}). 
The high correlation between signal and overall quality (0.98 at the model level and 0.93 at the clip level) can be attributed to the preponderance of signal impairments in this dataset, as opposed to other datasets such as DNS Challenges where background noise was the focus of the challenge. A majority (82\%) of the clips in this dataset were found to have lower signal quality than background noise (SIG $<$ BAK), whereas this number was below 30\% in the last DNS challenges. Given that the minimum of signal and background quality is a strong determinant of perceived overall quality~\cite{noauthor_itu-t_2003}, the observed high correlation between signal and overall quality in this dataset was expected.

\subsection{Correlation between P.804 and P.835}
The correlation between quality scores collected using P.804- and P.835-based subjective tests, for all entries are reported in Table~\ref{tab:sub-p804-835}. We observed a strong correlation between all the shared scores between the two subjective methodologies. Considering the rankings of participating teams, only the rank of N\&B and Kuaishou teams from the real-time track would swap when scores from P.835 test are used (tied rank using P.804 ratings).

\begin{table}
\centering
\caption{ Correlations between subjective scores obtained from P.804 and P.835 subjective tests on shared dimensions in model level for all entries. Tau-b95 is Kendall Tau-b applied to corrected ranked-order by considering 95\% confidence interval of subjective scores according to \cite{naderi_transformation_2020}.} 
\label{tab:sub-p804-835}
\begin{tabular}{l c c c c} 
\toprule
Dimension & PCC & SRCC & Kendall Tau-b & Tau-b95 \\
\midrule
Background/Noisiness & 0.964 & 0.926 & 0.825 & 0.853 \\
Signal & 0.954 & 0.933  & 0.801 & 0.914 \\
Overall & 0.965 & 0.940 & 0.825 & 0.822 \\
\midrule
M (challenge metric) & 0.961 & 0.946 & 0.825 & - \\
\bottomrule
\end{tabular}
\end{table}

\subsection{Correlation of subjective and objective data}
In Table \ref{fig:objective_results} we present the objective results on the blind set using DNSMOS  \cite{reddy_dnsmos_2022} (MOS$\_$SIG, MOS$\_$BAK, MOS$\_$OVR), and NISQA model \cite{mittag_nisqa_2021} (NISQA$\_$MOS, etc.). Similar to the subjective results, the Legends-tencent, Ctv-tencent, and Genius-team teams attained the best metrics estimated with DNSMOS and NISQA. Moreover, in Table \ref{tab:sub-obj-metrics} we compute the PCC between the subjective P.804 metrics and the metrics obtained with DNSMOS \cite{reddy_dnsmos_2022} and NISQA \cite{mittag_nisqa_2021}. The correlations range from PCC $0.478$ to $0.700$, which demonstrates why we still require a subjective test for accurately evaluating speech quality.

\begin{table}
\centering
\caption{ The PCC between the subjective P.804 results and the objective metrics estimated with DNSMOS P.835 \cite{reddy_dnsmos_2022} and NISQA \cite{mittag_nisqa_2021} models.} 
\setlength{\tabcolsep}{3pt}
\label{tab:sub-obj-metrics}
\begin{tabular}{l l c c} 
\toprule
Subjective metric & Objective metric & \multicolumn{2}{c}{PCC} \\
& & \small{Clip level} & \small{Model level}\\
\midrule
P.804 Overall & DNSMOS P.835 OVRL & 0.695 & 0.884\\
P.804 Overall & NISQA MOS & 0.681 & 0.766\\
P.804 Signal & DNSMOS P.835 SIG & 0.656 & 0.799\\
P.804 Noisiness & DNSMOS P.835 BAK & 0.545 & 0.933\\
P.804 Noisiness & NISQA NOISE & 0.586 & 0.938\\
P.804 Coloration & NISQA COLOR & 0.663 & 0.872\\
P.804 Discontinuity & NISQA DISCONTINUITY & 0.478 & 0.310\\
P.804 Loudness & NISQA LOUDNESS & 0.700 & 0.784\\
\bottomrule
\end{tabular}

\end{table}

\subsection{Model of Overall and other dimensions}
\label{sec:model}
We performed Explanatory Factory Analysis (EFA)~\cite{watkins_exploratory_2018} to investigate the underlying structure between the quality dimensions, namely if there is a shared variance between the sub-dimensions. We used the Maximum Likelihood extraction method with Varimax rotation and extracted three factors as suggested by the Scree plot~\cite{cattell_scree_1966}. 
The result of Bartlett’s test of sphericity was significant and the KMO value was 0.65 indicating that the data is adequate for explanatory factor analysis. 
The loading of quality scores on each factor is presented in Table~\ref{tab:factor_loading}. In total 62\% of the variance in data is explained by the three factors. Factor 1 represents the coloration with high loading from signal, coloration, and loudness. Discontinuity is loaded on factor 2 with some cross-loading from the signal indicating no or limited shared variance between discontinuity ratings and both coloration and loudness. As expected, noisiness built a separate factor orthogonal to others with moderate loading from reverberation. All in all, the results of EFA show that coloration, discontinuity, and noisiness are loaded on different orthogonal factors that align with the literature~\cite{waltermann_dimension-based_2013}. Signal scores share variance with coloration, discontinuity, and loudness, whereas reverberation shares variance with noisiness. 
Note that this factor structure represents the construct of the current training set and its generalizability should be validated in a separate study.

\begin{table}
\centering
\caption{The loading of quality scores on three-factor structure using Maximum Likelihood extraction method with Varimax rotation. KMO value = 0.65. Factor loading $>0.3$ is presented.} 
\label{tab:factor_loading}
\begin{tabular}{l c c c c} 
\toprule
Quality score & Factor 1 & Factor 2 & Factor 3 \\
\midrule
Signal & 0.824 & 0.481 & \\
Noisiness & & & 0.742 \\
Coloration & 0.787 & & \\
Discontinuity & & 0.936 & \\
Loudness & 0.476 & & \\
Reverberation & & & 0.413 \\
\bottomrule
\end{tabular}
\end{table}

In addition, we used different regressors to predict the overall quality given the subjective scores of the six sub-dimensions per clip. The results of k-fold cross-validations for clip and model levels are reported in Table~\ref{tab:model}.  Given that only a limited number of models are available in the dataset, random forest performed poorly compared to other regressors at the model level. The coefficients of the linear regression model and the feature importance from the random forest model are reported in Table~\ref{tab:importance}. At the clip level, the importance of features mostly agrees with both models.
Given the fact that most of the sub-dimensions have cross-loading with the signal quality in the explanatory factor analyses, we created different regressors to predict that. The performance of those regressors is reported in Table~\ref{tab:model_pred_sig} and the coefficients in Table~\ref{tab:importance_sig}. As expected, noisiness and reverberation have the smallest coefficients. 

\begin{table*}
\centering
\caption{Average performance of different regressors predicting overall quality given the six sub-dimensions in 5-fold cross-validation. In the model, level 3-fold cross-validation is used.} 
\label{tab:model}
\begin{tabular}{l c c c  c c c} 
\toprule
& \multicolumn{3}{c}{Clip level} & \multicolumn{3}{c}{Model level} \\
Regressor & PCC &  RMSE & $R^2$ & PCC &  RMSE & $R^2$ \\
\midrule
Linear regression & 0.947 & 0.318 & 0.894    & 0.993 & 0.051 & 0.951 \\ 
Polynomial (n=4)  & 0.959 & 0.276 & 0.920    & 0.996 & 0.047 & 0.969 \\ 
Random forest     & 0.960 & 0.276 & 0.921    & 0.977 & 0.131 & 0.754 \\ 
\bottomrule
\end{tabular}
\end{table*}

\begin{table*}
\centering
\caption{Average coefficient and importance of features in linear regression and random forest models predicting overall Quality, respectively.} 
\label{tab:importance}
\begin{tabular}{l c c c c } 
\toprule
& \multicolumn{2}{c}{Clip level} & \multicolumn{2}{c}{Model level} \\
Feature  & Linear regression & Random forest & Linear regression & Random forest\\
& coefficient & features imp. & coefficient & features imp. \\ 
\midrule
Signal & 0.646 &  0.878         & 0.352 & 0.378\\
Loudness & 0.146 &  0.044       & 0.251 & 0.162\\
Coloration & 0.102 &  0.019     & 0.266 & 0.146\\
Noisiness & 0.100 &  0.027      & 0.134 & 0.248\\
Discontinuity & 0.065 & 0.016   & 0.190 & 0.051\\
Reverberation & 0.039 & 0.016   & 0.014 & 0.016\\
\bottomrule
\end{tabular}
\end{table*}

\begin{table*}
\centering
\caption{Average performance of different regressors predicting signal quality given the five sub-dimensions in 5-fold cross-validation. In the model, level 3-fold cross-validation is used.} 
\label{tab:model_pred_sig}
\begin{tabular}{l c c c  c c c} 
\toprule
& \multicolumn{3}{c}{Clip level} & \multicolumn{3}{c}{Model level} \\
Regressor & PCC &  RMSE & $R^2$ & PCC &  RMSE & $R^2$ \\
\midrule
Linear regression & 0.898 & 0.453 & 0.806    & 0.994 & 0.035 & 0.979 \\ 
Polynomial (n=4)  & 0.907 & 0.434 & 0.821    & 0.992 & 0.043 & 0.964 \\ 
Random forest     & 0.901 & 0.446 & 0.812    & 0.852 & 0.170 & 0.452 \\ 
\bottomrule
\end{tabular}
\end{table*}

\begin{table*}
\centering
\caption{Average coefficient and importance of features in linear regression and random forest models predicting signal quality, respectively.} 
\label{tab:importance_sig}
\begin{tabular}{l c c c c } 
\toprule
& \multicolumn{2}{c}{Clip level} & \multicolumn{2}{c}{Model level} \\
Feature  & Linear regression & Random forest & Linear regression & Random forest\\
& coefficient & features imp. & coefficient & features imp. \\ 
\midrule
Loudness &      0.128 &  0.061      & 0.184 & 0.142\\
Coloration &    0.503 &  0.559      & 0.519 & 0.373\\
Noisiness &     0.072 &  0.044      & 0.051 & 0.306\\
Discontinuity & 0.430 &  0.281      & 0.580 & 0.130\\
Reverberation & 0.099 &  0.054      & 0.158 & 0.049\\
\bottomrule
\end{tabular}
\end{table*}


\subsection{Word error rate}
\begin{table*}
    \centering
    \caption{Real-time track word error rate challenge results for the blind test set.}
    \includegraphics[width=\textwidth]{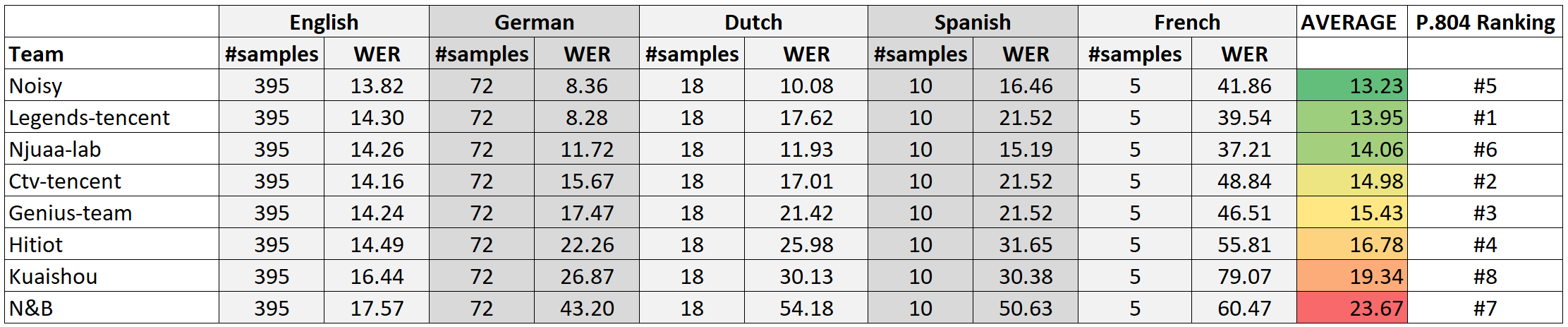}
    \label{fig:track1_wer_results}
\end{table*}

\begin{table*}
    \centering
    \caption{Non-real-time track word error rate challenge results for the blind test set. The teams marked with * have identical submissions for both tracks.}
    \includegraphics[width=\textwidth]{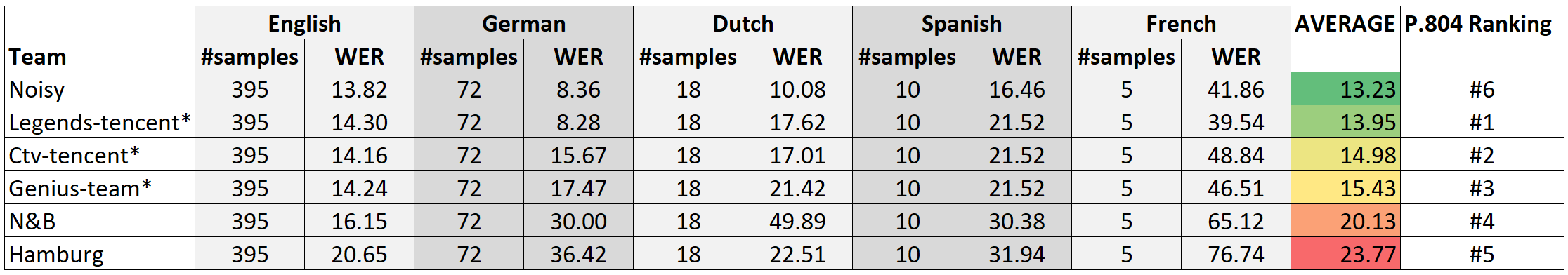}
    \label{fig:track2_wer_results}
\end{table*}

To have a more comprehensive view of the signal enhancement models, in Table \ref{fig:track1_wer_results} and Table \ref{fig:track2_wer_results} we included the word error rate (WER) for both tracks. To eliminate potential bias introduced by automatic speech recognition (ASR) systems, we employed human transcripts when calculating the WER. A state-of-the-art speech recognition API from Azure
Cognitive service was used for computing WER. In the second track, the rank is identical to the P.804 ranking (excluding noisy), while in the first track, there are some shifts between teams. The best WER result attained by Legends-tencent team is still slightly behind the WER computed on the noisy files, highlighting that there is a huge potential in this research area.  

\section{Conclusions}
Unlike our previous deep noise suppression challenges, this challenge showed several models with significant improvement in the speech signal. The top models improved all areas we measured: noisiness, discontinuity, coloration, loudness, and reverberation. While the improvements are impressive, there is still significant room for improvement in this test set (see Table \ref{tab:improvement2}).

All of the models used in this challenge are relatively small compared to large language models or large multimodal language models. An interesting new area would be to apply a large audio language model (e.g., \cite{borsos_audiolm_2023}) for speech restoration and enhancement. Even if it can not be run in real-time or with low latency, there are still many scenarios it can be applied.
In addition, all of the models submitted in this challenge used training sets with clean speech available. A good future direction of research is to utilize real-world training sets such as \cite{cutler_icassp_2022}, which will require semi-supervised or unsupervised learning. 

\begin{table}
    \centering
    \caption{Amount of improvement remaining (in MOS) to get excellent quality rated speech based on this challenge}
    \begin{tabular}{l c}
    \toprule
        Area & Headroom \\
       \midrule
        Overall & 1.73 \\
        Signal & 1.74 \\
        Background & 0.36 \\
        Coloration & 1.40 \\
        Loudness &  0.80 \\
        Discontinuity & 0.89 \\
        Reverberation & 0.68 \\
        \bottomrule
    \end{tabular}
    \label{tab:improvement2}
\end{table}

For future speech signal improvement challenges, we plan to provide an objective metric similar to NISQA. We plan to also add word accuracy rate as an additional metric to optimize. We plan to provide a synthetic data generator and a baseline model to give a better starting point for all participants. As noted above, we hypothesize that large multimodal models could have significant improvements in this area, so keeping a non-real-time track seems important to encourage this exploration. 

\FloatBarrier 
\bibliographystyle{IEEEbib}
\bibliography{IC3-AI}

\begin{thebibliography}{10}

\bibitem{noauthor_itu-t_2003}
``{ITU}-{T} {Recommendation} {P}.835: {Subjective} test methodology for
  evaluating speech communication systems that include noise suppression
  algorithm,'' 2003.

\bibitem{dubey_icassp_2022}
H.~Dubey, V.~Gopal, R.~Cutler, A.~Aazami, S.~Matusevych, S.~Braun, S.~E.
  Eskimez, M.~Thakker, T.~Yoshioka, H.~Gamper, and R.~Aichner,
\newblock ``{ICASSP} 2022 {Deep} {Noise} {Suppression} {Challenge},''
\newblock in {\em {ICASSP}}, 2022.

\bibitem{reddy_interspeech_2021}
C.~K. Reddy, H.~Dubey, K.~Koishida, A.~Nair, V.~Gopal, R.~Cutler, S.~Braun,
  H.~Gamper, R.~Aichner, and S.~Srinivasan,
\newblock ``{INTERSPEECH} 2021 {Deep} {Noise} {Suppression} {Challenge},''
\newblock in {\em {INTERSPEECH}}. Aug. 2021, pp. 2796--2800, ISCA.

\bibitem{dubey_icassp_2023}
H.~Dubey, A.~Aazami, V.~Gopal, B.~Naderi, S.~Braun, R.~Cutler, A.~Ju,
  M.~Zohourian, M.~Tang, H.~Gamper, M.~Golestaneh, and R.~Aichner,
\newblock ``{ICASSP} 2023 {Deep} {Noise} {Suppression} {Challenge},'' May 2023,
\newblock arXiv:2303.11510 [cs, eess].

\bibitem{noauthor_itu-t_2017-1}
``{ITU}-{T} {Recommendation} {P}.804: {Subjective} diagnostic test method for
  conversational speech quality analysis,'' 2017.

\bibitem{su_hifi-gan-2_2021}
J.~Su, Z.~Jin, and A.~Finkelstein,
\newblock ``{HiFi}-{GAN}-2: {Studio}-{Quality} {Speech} {Enhancement} via
  {Generative} {Adversarial} {Networks} {Conditioned} on {Acoustic}
  {Features},''
\newblock in {\em 2021 {IEEE} {Workshop} on {Applications} of {Signal}
  {Processing} to {Audio} and {Acoustics} ({WASPAA})}, Oct. 2021, pp. 166--170,
\newblock ISSN: 1947-1629.

\bibitem{serra_universal_2022}
J.~Serrà, S.~Pascual, J.~Pons, R.~O. Araz, and D.~Scaini,
\newblock ``Universal {Speech} {Enhancement} with {Score}-based {Diffusion},''
  2022,
\newblock https://openreview.net/forum?id=7BfWbjOqgMf.

\bibitem{noauthor_itu-t_1996}
``{ITU}-{T} {Recommendation} {P}.800: {Methods} for subjective determination of
  transmission quality,'' 1996.

\bibitem{noauthor_itu-t_2018}
``{ITU}-{T} {Recomendation} {P}.808: {Subjective} evaluation of speech quality
  with a crowdsourcing approach,'' 2018.

\bibitem{naderi_open_2020}
B.~Naderi and R.~Cutler,
\newblock ``An {Open} {Source} {Implementation} of {ITU}-{T} {Recommendation}
  {P}.808 with {Validation},''
\newblock {\em INTERSPEECH}, pp. 2862--2866, Oct. 2020.

\bibitem{naderi_subjective_2021}
B.~Naderi and R.~Cutler,
\newblock ``Subjective {Evaluation} of {Noise} {Suppression} {Algorithms} in
  {Crowdsourcing},''
\newblock in {\em {INTERSPEECH}}, 2021.

\bibitem{noauthor_itu-t_2022}
``{ITU}-{T} {Recommendation} {P}.863.2: {Extension} of {ITU}-{T} {P}.863 for
  multi-dimensional assessment of degradations in telephony speech signals up
  to full-band,'' 2022.

\bibitem{rix_perceptual_2001}
A.~Rix, J.~Beerends, M.~Hollier, and A.~Hekstra,
\newblock ``Perceptual evaluation of speech quality ({PESQ}) - a new method for
  speech quality assessment of telephone networks and codecs,''
\newblock in {\em {ICASSP}}. 2001, vol.~2, pp. 749--752, IEEE.

\bibitem{beerends_perceptual_2013}
J.~G. Beerends, M.~Obermann, R.~Ullmann, J.~Pomy, and M.~Keyhl,
\newblock ``Perceptual {Objective} {Listening} {Quality} {Assessment}
  ({POLQA}), {The} {Third} {Generation} {ITU}-{T} {Standard} for {End}-to-{End}
  {Speech} {Quality} {Measurement} {Part} {I}–{Temporal} {Alignment},''
\newblock {\em J. Audio Eng. Soc.}, vol. 61, no. 6, pp. 19, 2013.

\bibitem{noauthor_itu-t_2011}
``{ITU}-{T} {Recommendation} {P}.563: {Perceptual} objective listening quality
  assessment: {An} advanced objective perceptual method for end-to-end
  listening speech quality evaluation of fixed, mobile, and {IP}-based networks
  and speech codecs covering narrowband, wideband, and super-wideband
  signals,'' 2011.

\bibitem{avila_non-intrusive_2019}
A.~R. Avila, H.~Gamper, C.~Reddy, R.~Cutler, I.~Tashev, and J.~Gehrke,
\newblock ``Non-intrusive {Speech} {Quality} {Assessment} {Using} {Neural}
  {Networks},''
\newblock in {\em {ICASSP}}, Brighton, United Kingdom, May 2019, pp. 631--635,
  IEEE.

\bibitem{reddy_dnsmos_2021}
C.~K.~A. Reddy, V.~Gopal, and R.~Cutler,
\newblock ``{DNSMOS}: {A} {Non}-{Intrusive} {Perceptual} {Objective} {Speech}
  {Quality} metric to evaluate {Noise} {Suppressors},''
\newblock in {\em {INTERSPEECH}}, 2021.

\bibitem{reddy_dnsmos_2022}
C.~K.~A. Reddy, V.~Gopal, and R.~Cutler,
\newblock ``{DNSMOS} {P}.835: {A} {Non}-{Intrusive} {Perceptual} {Objective}
  {Speech} {Quality} {Metric} to {Evaluate} {Noise} {Suppressors},''
\newblock in {\em {ICASSP}}, 2022, pp. 886--890,
\newblock ISSN: 2379-190X.

\bibitem{yi_conferencingspeech_2022}
G.~Yi, W.~Xiao, Y.~Xiao, B.~Naderi, S.~Möller, W.~Wardah, G.~Mittag,
  R.~Cutler, Z.~Zhang, D.~S. Williamson, F.~Chen, F.~Yang, and S.~Shang,
\newblock ``{ConferencingSpeech} 2022 {Challenge}: {Non}-intrusive {Objective}
  {Speech} {Quality} {Assessment} ({NISQA}) {Challenge} for {Online}
  {Conferencing} {Applications},''
\newblock in {\em {INTERSPEECH}}, 2022.

\bibitem{mittag_nisqa_2021}
G.~Mittag, B.~Naderi, A.~Chehadi, and S.~Möller,
\newblock ``{NISQA}: {A} deep cnn-self-attention model for multidimensional
  speech quality prediction with crowdsourced datasets,''
\newblock in {\em {INTERSPEECH}}, 2021.

\bibitem{reddy_interspeech_2020}
C.~K. Reddy, V.~Gopal, R.~Cutler, E.~Beyrami, R.~Cheng, H.~Dubey,
  S.~Matusevych, R.~Aichner, A.~Aazami, S.~Braun, P.~Rana, S.~Srinivasan, and
  J.~Gehrke,
\newblock ``The {INTERSPEECH} 2020 {Deep} {Noise} {Suppression} {Challenge}:
  {Datasets}, {Subjective} {Testing} {Framework}, and {Challenge} {Results},''
\newblock in {\em {INTERSPEECH} 2020}. Oct. 2020, pp. 2492--2496, ISCA.

\bibitem{reddy_icassp_2021}
C.~K.~A. Reddy, H.~Dubey, V.~Gopal, R.~Cutler, S.~Braun, H.~Gamper, R.~Aichner,
  and S.~Srinivasan,
\newblock ``{ICASSP} 2021 {Deep} {Noise} {Suppression} {Challenge},''
\newblock in {\em {ICASSP}}, June 2021, pp. 6623--6627,
\newblock ISSN: 2379-190X.

\bibitem{diener_interspeech_2022}
L.~Diener, S.~Sootla, S.~Branets, A.~Saabas, R.~Aichner, and R.~Cutler,
\newblock ``{INTERSPEECH} 2022 {Audio} {Deep} {Packet} {Loss} {Concealment}
  {Challenge},''
\newblock in {\em {INTERSPEECH}}, 2022, p.~5.

\bibitem{kinoshita_summary_2016}
K.~Kinoshita, M.~Delcroix, S.~Gannot, E.~A. P.~Habets, R.~Haeb-Umbach,
  W.~Kellermann, V.~Leutnant, R.~Maas, T.~Nakatani, B.~Raj, A.~Sehr, and
  T.~Yoshioka,
\newblock ``A summary of the {REVERB} challenge: state-of-the-art and remaining
  challenges in reverberant speech processing research,''
\newblock {\em EURASIP Journal on Advances in Signal Processing}, vol. 2016,
  no. 1, pp. 7, Dec. 2016.

\bibitem{sridhar_icassp_2021}
K.~Sridhar, R.~Cutler, A.~Saabas, T.~Parnamaa, M.~Loide, H.~Gamper, S.~Braun,
  R.~Aichner, and S.~Srinivasan,
\newblock ``{ICASSP} 2021 {Acoustic} {Echo} {Cancellation} {Challenge}:
  {Datasets}, {Testing} {Framework}, and {Results},''
\newblock in {\em {ICASSP}}, 2021.

\bibitem{cutler_interspeech_2021}
R.~Cutler, A.~Saabas, T.~Parnamaa, M.~Loide, S.~Sootla, M.~Purin, H.~Gamper,
  S.~Braun, K.~Sorensen, R.~Aichner, and S.~Srinivasan,
\newblock ``{INTERSPEECH} 2021 {Acoustic} {Echo} {Cancellation} {Challenge},''
\newblock in {\em {INTERSPEECH}}, June 2021.

\bibitem{cutler_icassp_2022}
R.~Cutler, A.~Saabas, T.~Parnamaa, M.~Purin, H.~Gamper, S.~Braun, and
  R.~Aichner,
\newblock ``{ICASSP} 2022 {Acoustic} {Echo} {Cancellation} {Challenge},''
\newblock in {\em {ICASSP}}, 2022.

\bibitem{cutler_icassp_2023}
R.~Cutler, A.~Saabas, T.~Parnamaa, M.~Purin, E.~Indenbom, N.-C. Ristea,
  J.~Gužvin, H.~Gamper, S.~Braun, and R.~Aichner,
\newblock ``{ICASSP} 2023 {Acoustic} {Echo} {Cancellation} {Challenge},''
\newblock 2023.

\bibitem{li_dds_2022}
H.~Li and J.~Yamagishi,
\newblock ``{DDS}: {A} new device-degraded speech dataset for speech
  enhancement,''
\newblock in {\em {INtERSPEECH}}, 2022.

\bibitem{mysore_can_2015}
G.~J. Mysore,
\newblock ``Can we {Automatically} {Transform} {Speech} {Recorded} on {Common}
  {Consumer} {Devices} in {Real}-{World} {Environments} into {Professional}
  {Production} {Quality} {Speech}?—{A} {Dataset}, {Insights}, and
  {Challenges},''
\newblock {\em IEEE Signal Processing Letters}, vol. 22, no. 8, pp. 1006--1010,
  Aug. 2015,
\newblock Conference Name: IEEE Signal Processing Letters.

\bibitem{naderi_multi-dimensional_2023}
B.~Naderi, R.~Cutler, and N.-C. Ristea,
\newblock ``Multi-dimensional {Speech} {Quality} {Assessment} in
  {Crowdsourcing},'' Sept. 2023,
\newblock arXiv:2309.07385 [cs, eess].

\bibitem{naderi_towards_2020}
B.~Naderi, R.~Zequeira~Jiménez, M.~Hirth, S.~Möller, F.~Metzger, and
  T.~Hoßfeld,
\newblock ``Towards speech quality assessment using a crowdsourcing approach:
  evaluation of standardized methods,''
\newblock {\em Quality and User Experience}, vol. 6, no. 1, pp. 2, Nov. 2020.

\bibitem{naderi_application_2020}
B.~Naderi and S.~Möller,
\newblock ``Application of just-noticeable difference in quality as environment
  suitability test for crowdsourcing speech quality assessment task,''
\newblock in {\em {QoMEX}}. 2020, pp. 1--6, IEEE.

\bibitem{naderi_effect_2015}
B.~Naderi, T.~Polzehl, I.~Wechsung, F.~Köster, and S.~Möller,
\newblock ``Effect of trapping questions on the reliability of speech quality
  judgments in a crowdsourcing paradigm,''
\newblock in {\em {INTERSPEECH}}, 2015.

\bibitem{chen_gesper_2023}
J.~Chen, Y.~Shi, W.~Liu, W.~Rao, S.~He, A.~Li, Y.~Wang, Z.~Wu, S.~Shang, and
  C.~Zheng,
\newblock ``Gesper: {A} unified framework for general speech restoration,''
\newblock in {\em {ICASSP}}, 2023.

\bibitem{zhu_ssi-net_2023}
W.~Zhu, Z.~Wang, J.~Lin, C.~Zeng, and T.~Yu,
\newblock ``{SSI}-{NET}: {A} multi-stage speech signal improvement system for
  {ICASSP} 2023 {SSI} challenge,''
\newblock in {\em {ICASSP}}, 2023.

\bibitem{zhang_half-temporal_2023}
Z.~Zhang, S.~Xu, X.~Zhuang, Y.~Qian, L.~Zhou, and M.~Wang,
\newblock ``Half-temporal and half-frequency attention u2net for speech signal
  improvement,''
\newblock in {\em {ICASSP}}, 2023.

\bibitem{liu_two-stage_2023}
M.~Liu, S.~Lv, Z.~Zhang, R.~Han, X.~Hao, X.~Xia, L.~Chen, Y.~Xiao, and L.~Xie,
\newblock ``Two-stage neural network for {ICASSP} 2023 speech signal
  improvement challenge,''
\newblock in {\em {ICASSP}}, 2023.

\bibitem{richter_speech_2023}
J.~Richter, S.~Welker, J.-M. Lemercier, B.~Lay, T.~Peer, and T.~Gerkmann,
\newblock ``Speech signal improvement using causal generative diffusion
  models,''
\newblock in {\em {ICASSP}}, 2023.

\bibitem{yamagishi_cstr_2019}
J.~Yamagishi, C.~Veaux, and K.~MacDonald,
\newblock ``{CSTR} {VCTK} {Corpus}: {English} multi-speaker corpus for {CSTR}
  voice cloning toolkit (version 0.92),'' 2019,
\newblock https://doi.org/10.7488/ds/2645.

\bibitem{naderi_transformation_2020}
B.~Naderi and S.~Möller,
\newblock ``Transformation of {Mean} {Opinion} {Scores} to {Avoid} {Misleading}
  of {Ranked} {Based} {Statistical} {Techniques},''
\newblock in {\em {QoMEX}}, May 2020, pp. 1--4,
\newblock ISSN: 2472-7814.

\bibitem{watkins_exploratory_2018}
M.~W. Watkins,
\newblock ``Exploratory {Factor} {Analysis}: {A} {Guide} to {Best}
  {Practice},''
\newblock {\em Journal of Black Psychology}, vol. 44, no. 3, pp. 219--246, Apr.
  2018,
\newblock Publisher: SAGE Publications Inc.

\bibitem{cattell_scree_1966}
R.~B. Cattell,
\newblock ``The {Scree} {Test} {For} {The} {Number} {Of} {Factors},''
\newblock {\em Multivariate Behavioral Research}, vol. 1, no. 2, pp. 245--276,
  Apr. 1966,
\newblock Publisher: Routledge \_eprint:
  https://doi.org/10.1207/s15327906mbr0102\_10.

\bibitem{waltermann_dimension-based_2013}
M.~Wältermann,
\newblock {\em Dimension-based {Quality} {Modeling} of {Transmitted} {Speech}},
\newblock Springer Science \& Business Media, Jan. 2013,
\newblock Google-Books-ID: \_TJEAAAAQBAJ.

\bibitem{borsos_audiolm_2023}
Z.~Borsos, R.~Marinier, D.~Vincent, E.~Kharitonov, O.~Pietquin, M.~Sharifi,
  D.~Roblek, O.~Teboul, D.~Grangier, M.~Tagliasacchi, and N.~Zeghidour,
\newblock ``{AudioLM}: {A} {Language} {Modeling} {Approach} to {Audio}
  {Generation},''
\newblock {\em IEEE/ACM Transactions on Audio, Speech, and Language
  Processing}, vol. 31, pp. 2523--2533, 2023,
\newblock Conference Name: IEEE/ACM Transactions on Audio, Speech, and Language
  Processing.

\end{thebibliography}

\end{document}